\documentstyle[mprocl,psfig]{article}

\def\PRD{{\it Phys. Rev.} D}

\def\CQG{{\it Class. Quantum Gravity} }

\def\IJMP{{\it Int. J. Mod. Phys.} }

\def\NP{{\it Nucl. Phys.} }

\def\PR{{\it Phys. Rev.} }

\def\beq{\begin{equation}}
\def\eeq{\end{equation}}

\def\be{\begin{equation}}
\def\ee{\end{equation}}

\def\bea{\begin{eqnarray}}
\def\eea{\end{eqnarray}}

\begin{document}

\title{QUANTUM ANALYSIS OF THE COMPACTIFICATION PROCESS IN THE 
MULTIDIMENSIONAL EINSTEIN-YANG-MILLS SYSTEM}

\author{ O. BERTOLAMI \footnote{Speaker} }

\address{Departamento de F\'\i sica, Instituto Superior T\'ecnico,\\
Av. Rovisco Pais, Lisbon, Portugal}

\author{ P.D. FONSECA }

\address{Department of Physics and Astronomy, Rutgers University,\\
Piscataway, NJ 08855-0849, U.S.A.}

\author { P.V. MONIZ }

\address{University of Cambridge, DAMTP, Silver Street,\\
Cambridge CB3 9EW, England}

\maketitle\abstracts{We study solutions of the Wheeler-DeWitt equation 
obtained when considering homogeneous and isotropic 
(up to a gauge transformation) field configurations of the Einstein-Yang-Mills
system in $D=4+d$ dimensions with an ${\bf R} \times S^3 \times S^d$ topology
and assuming the Hartle-Hawking boundary conditions.}

\section{Introduction}

We report in this contribution on the result of recent work on the quantum 
cosmology of the minisuperspace model arising from 
the coset space dimensional reduction of the 
$D$-dimensional Einstein-Yang-Mills (EYM) sytem \cite{bfm}.
Our method consists in exploiting the
isometries of an homogeneous and isotropic spacetime in four 
\cite{bmpv,bm,bmon} 
or $D$-dimensions \cite{bkm} to restrict the possible field configurations. 
We find
that compactifying solutions correspond to maxima of the wave function
indicating that these solutions are favoured over the ones where the
extra dimensions are not compactified for an expanding Universe. We
also find that some features of the wave function of the Universe do
depend on the number of extra dimensions \cite{bfm}. 

\section{Effective Model and Wheeler-DeWitt equation}

We consider a $D=d+4$-dimensional EYM model
restricted to spatially homogeneous and (partially) isotropic field
configurations, i.e. symmetric fields (up to gauge transformations for
the gauge field) under the action of a group $G^{{\rm ext }} \times
G^{{\rm int}}$ and the gauge group $\hat K = SO(N), N \geq 3+d$.
 Introducing the most general $SO(4) \times SO(d+1)$-invariant metric in 
the space $M^D = M^{4 + d} = M^4 \times I^d$ 
and the $SO(4) \times SO(d+1)-$ symmetric Ansatz for the
gauge field \cite{bkm} (see also Ref. [2] for a general discussion)
into the $D$-dimensional EYM action leads to the following 
effective model \cite{bkm}:
\bea
S _{\rm eff} = 
 16 \pi^2 \int dt N a^3  \Biggl\{ -{3 \over 8\pi k a^{2}} 
\left(\left[{\dot{a} \over N}\right]^2-{1\over 4}\right)   
+ {1 \over 2} \left[{\dot{\psi} \over N}\right]^2
 +  e^{d\beta\psi} {3 \over 4e^2 a^{2}} 
\left( {1 \over 2}\left[{\dot{f_0} \over N}\right]^2 \right.\nonumber \\
+  \left.{1 \over 2}\left[{ {\cal D}_t {\bf f} \over N}\right]^2\right) 
+  e^{-2\beta\psi} {d \over 8e^2 b_0^{2}}
\left[{ {\cal D}_t {\bf g} \over N}\right]^2 
 -   W(a,\psi,f_0,{\bf f}, 
{\bf g}) \Biggr\}~, 
\label{eq:4.5}
\eea
where $k = \hat k /v_d b_0^d, e^2 = \hat e^2/v_d b_0^d, 
\beta = \sqrt{16 \pi k / d(d+2)}, v_d$ is the the volume 
of $S^d$ for $b=1$, 
$\psi = \beta^{-1} \ln (b/b_0)$, $b_0$ being the equilibrium value of the 
internal space scale factor, $b$. Moreover, 
${\cal D}_t$ denotes the covariant derivative with 
respect to the remnant $SO(N-3-d)$ gauge field $\hat B(t)$: 
${\cal D}_t {\bf f}(t) = {d \over d t} {\bf f(t)} + 
\hat B(t) {\bf f}(t)$, and
${\cal D}_t {\bf g}(t) = {d \over d t} {\bf g(t)} + 
\hat B(t) {\bf g}(t)$, such that ${\bf f} = \left\{ f_p \right\}$,  
${\bf g} = \left\{ g_q \right\}$  
and $\hat B$ is an
$(N-3-d)\times (N-3-d)$ antisymmetric matrix 
$\hat B = (B_{pq})$. The potential $W$ is on its hand given by: 
\begin{eqnarray}
W & = & e^{-d\beta\psi} \left[ -e^{-2\beta\psi}
{d(d-1) \over 64\pi k b_0^{2}} \right.
+ \left. e^{-4\beta\psi} {d(d-1) \over 8 e^2 b_0^{4}} V_2 ({\bf g})
+  {\Lambda \over 8\pi k}\right] \nonumber \\
& + & e^{-2\beta\psi} {3d \over 32 e^2 (ab_0)^{2}} 
({\bf f}\cdot {\bf g})^2
+ e^{d\beta\psi} {3 \over 4 e^2 a^4} V_1(f_0, {\bf f})~,
\end{eqnarray}
where $\Lambda = v_d b_0^d \hat \Lambda$, and 
$V_1 (f_0, {\bf f})  =   {1 \over 8} 
\left[ \left( f_0^2 + {\bf f}^2 - 1 \right)^2 + 
4 f_0^2 {\bf f}^2 \right]$,
$V_2 ({\bf g})  =  {1 \over 8} \left( {\bf g}^2  - 1\right)^2$
are the contributions associated with the external and internal components 
of the gauge field, respectively.

Introducing the new variables $(\mu, \phi)$
$$
a =  \left({k \over 6\pi}\right)^{\frac{1}{2}} e^\mu,
~\indent
\psi =  \left({3 \over 4\pi k }\right)^{\frac{1}{2}} \phi ~,
$$
as well as $\epsilon \equiv \sqrt{12/d (d+2)}$ 
and turning the
canonical conjugate momenta into operators, $ \pi_\mu \mapsto
-i\frac{\partial}{\partial \mu}$, $\pi_\phi \mapsto
-i\frac{\partial}{\partial \phi}$, etc., leads to the 
Wheeler-DeWitt equation for the wave function $\Psi=\Psi(\mu, \phi,
f_0, {\bf f}, {\bf g})$ \cite{bfm}. 
To study the compactification process we further set 
$~f_0 = f_0^v, ~{\bf f} = {\bf f}^v,
~{\bf g} = {\bf g}^v = {\bf 0}~ and 
{\bf f} \cdot {\bf g} = 0$. From the definitons
$v_1 \equiv V_1 (f_0^v, {\bf f}^v)$,  
$~v_2 \equiv V_2({\bf g}^v)={1 \over 8}~$ and, after setting 
$p=0$ when parametrizating
the operator order ambiguity $\pi_\mu^2 \mapsto - \mu^{-p}
\frac{\partial}{\partial \mu} \left(\mu^p \frac{\partial}{\partial
\mu} \right)$,
our problem simplifies to \cite{bfm}
\beq
\left[{\partial^2 \over \partial \mu^2} - {\partial^2 \over \partial
\phi^2} + U(\mu, \phi) \right] \Psi(\mu,\phi) = 0, 
\eeq 
where 
\beq
U(\mu, \phi) = e^{6\mu} \left({4 k \over 3}\right)^2 \Omega (\mu,
\phi) - e^{4 \mu}~, 
\eeq  
\beq 
\Omega (\mu, \phi)  = -e^{-(d+2)\epsilon\phi} 
{d(d-1) \over 64\pi k b_0^{2}}
+ e^{-(d+4)\epsilon\phi} v_2{d(d-1) \over 8 e^2 b_0^{4}}
+ e^{d\epsilon\phi -4 \mu} v_1{27 \over  e^2 k^2}
+ {\Lambda \over 8\pi k}.
\eeq

\section{Solutions with dynamical compactification}

Ensuring the stability of compactifation and satifying 
the observational bound,
$|\Lambda^{(4)}|<10^{-120}\frac{1}{16\pi k}$, requires 
the cosmological constant to fulfil the
condition \cite{bkm}:
\beq
\Lambda={d(d-1) \over 16b_0^2}~,
\eeq 
where $b_0^2 = 16\pi k
v_2/e^2$. 
The potential in the minisuperspace simplifies then to
\begin{eqnarray} 
U(\mu,\phi) = e^{6 \mu-d \epsilon \phi} {2k \Lambda
\over 9 \pi} \left( e^{-2\epsilon\phi}-1 \right) ^2-e^{4\mu}
+ e^{2\mu+d \epsilon \phi} {3\pi \over k} {v_1 \over v_2}b_0^2 ~.
\end{eqnarray}

Next we consider the Hartle-Hawking path integral representation for
the ground-state wave function of the Universe \cite{hh} 
\beq
\Psi[\mu,\phi]=\int_{C}{D\mu D\phi \exp(-S_{\rm E})}~, 
\eeq 
which allows to evaluate the solution of (49), $\Psi(\mu,\phi)$, close to
$\mu=-\infty$ ($S_{\rm E}=-iS_{\rm eff}$ and C is a compact manifold
with no boundary) and to establish the regions where the wave function
behaves as an exponential (quantum regime) or as an oscillation
(classical regime). Our results indicate that a generic feature of the wave
function is that solutions corresponding to stable compactifying
solutions are maxima, meaning that they are indeed the most probable 
configurations, for an
expanding Universe. Moreover, some
properties of the wave function were found to depend on the number,
$d$, of internal space dimensions, namely, the regions
where the wave function predicts the 4-dimensional metric behaves classically
or quantum mechanically (i.e. regions where the metric is Lorentzian
or Euclidean) will differ between the $d<4$ and the \mbox{$d\ge 4$} cases
\cite{bfm}.

We stress that a distinctive feature of our scheme is the
non-vanishing contribution of the external components of the gauge
field to the potential $V_1$ in $W$.  It is precisely this feature
that allows obtaining a classically stable
compactification after inflation \cite{bkm} and that is responsible
for some of the dependence of the wave function in the number, $d$, of
internal dimensions \cite{bfm}. In this respect our work differs from 
previous one where stable compactification is achieved through the internal
components of a magnetic monopole \cite{jjh} or via a $d$-rank antisymmetric
tensor \cite{cw}.


\section*{References}

\begin{thebibliography}{99}


\bibitem{bfm} O. Bertolami, P.D. Fonseca and P.V. Moniz, ``Quantum
Cosmological Multidimensional Einstein-Yang-Mills Model in a $R \times
S^3 \times S^{d}$ Topology'' (gr-qc 9607015), to appear in \PRD.


\bibitem{bmpv}O. Bertolami, J.M. Mour\~ao, R.F. Picken and
I.P. Volobujev, \IJMP {\bf A6} (1991) 4149.


\bibitem{bm}O. Bertolami and J.M. Mour\~ao, \CQG {\bf 8} (1991) 1271.


\bibitem{bmon}O. Bertolami and P.V. Moniz, \NP {\bf B439} (1995) 259.


\bibitem{bkm}O. Bertolami, Yu.A. Kubyshin and J.M. Mour\~ao, \PR {\bf
D45} (1992) 3405.


\bibitem{hh}J.B. Hartle and S.W. Hawking, \PR {\bf D28} (1983) 2960.

 
\bibitem{jjh}J.J. Halliwell, \NP {\bf B266} (1986) 228.


\bibitem{cw}U. Carow-Watamura, T. Inami and S. Watamura, \CQG {\bf 4}
(1987) 23.

\end{thebibliography}
\end{document}